\documentclass[11pt,a4paper]{JHEP}
\usepackage{epsfig}
\usepackage{psfrag}
\usepackage{amssymb}
\usepackage{amsmath}
\newcommand{\bq}{\begin{eqnarray}}
\newcommand{\eq}{\end{eqnarray}}
\newcommand{\bqs}{\begin{eqnarray*}}
\newcommand{\eqs}{\end{eqnarray*}}
\newcommand{\p}{\partial}

\newcommand{\IC}{\mathbb{C}}
\newcommand{\Psifr}{\Psi_{\text{fr}}}

\newcommand{\YTvert}{
\begin{array}{c}
\square\\[-1.7ex]
\square
\end{array}
}
\newcommand{\YThor}{
\square\!\square
}
\newlength{\mylength}
\newcommand{\kaderzonderrm}[1]
{\ \\[2ex]
  \setlength{\fboxsep}{1ex}
  \setlength{\mylength}{\linewidth}
  \addtolength{\mylength}{-2\fboxsep}
  \addtolength{\mylength}{-2\fboxrule}
  \framebox[\linewidth]{
    \begin{minipage}{\mylength}
      \vspace{1ex}
      #1
    \end{minipage}
  }\ \\[2ex]}
\def\a{\alpha}

\def\d{\delta}
\def\e{\varepsilon}
\def\vp{\varphi}
\def\g{\gamma}
\def\k{\kappa}   
\def\l{\lambda}
\def\m{\mu}
\def\n{\nu}
\def\o{\omega}                    
\def\th{\theta}                                   
\def\s{\sigma}                
\def\O{\Omega}
\title{Black holes on cylinders are not algebraically special}
\author{Pieter-Jan De Smet\footnote{Address after 1 
September 2002: SUNY at Stony Brook, USA}\\
Instituut voor theoretische fysica, Katholieke
Universiteit Leuven,\\
Celestijnenlaan 200D, B-3001 Leuven, Belgium. \\
E-mail: {\tt Pieter-Jan.DeSmet@fys.kuleuven.ac.be} }
\preprint{KUL-TF-2002/06 \\{\tt hep-th/0206106}}
\abstract{We give a Petrov classification for  five-dimensional metrics. 
We classify Ricci-flat metrics that are static,
have an $SO(3)$ isometry group and have Petrov type 22.
We use this classification to look for the metric of a black hole on 
a cylinder, i.e.~a black hole with asymptotic geometry four-dimensional
Minkowski space times a circle. Although a black string wrapped around the 
$S^1$ and  the five-dimensional black hole are both algebraically special, 
it turns out that the black hole on a cylinder is not.}
\begin{document}
\setcounter{footnote}{0}
\section{Introduction}
In this article, we try to construct the metric of a neutral black hole in a 
space with one compact dimension. More specifically, we look for a black hole
in a 5-dimensional spacetime which is asymptotically the product of
the 4-dimensional Minkowski space and a circle. Henceforth, we will refer
to this kind of black hole as \textit{a black hole on a cylinder}. 

The Newton potential of a point mass on a cylinder is easily found.
Indeed, because 
of the linearity of the Laplace operator, one can determine 
this potential by a summation over an infinite number of
mirror masses. However, the Einstein equations are non-linear and, hence,
determining the metric of a (neutral) black hole on a cylinder is much more
difficult\footnote{The metric of a maximally
charged black hole on a cylinder is known~\cite{Myers}. 
In this case, because there is no force
between extremal black holes, the non-linearities in the Einstein equations
are less severe and hence easier to solve. 
One can also consider this simplification as a consequence of supersymmetry.}. 
Its metric is presumably complicated because neither the spherical symmetry nor
the cylindrical symmetry applies in general. Therefore, the metric depends on
two coordinates -- on the radial distance in the large dimensions and on the
coordinate in the compact dimension -- and the Einstein equations lead to a
system of partial differential equations.

Instead of a direct attempt to integrate these complicated equations, our line 
of attack will be the following. First, we determine the Petrov type of
the 5-dimensional Schwarzschild black hole and of the black string wrapping the
compact circle. These two geometries can be viewed as limiting cases of
the black hole on the cylinder. Indeed, very close to the black hole on the
cylinder, the size of the compact circle is irrelevant and the metric approaches
the 5-dimensional Schwarzschild geometry. On the other hand, far from the
black hole on the cylinder, the fact that the black hole is localized is 
unimportant and the metric approaches the one of the wrapped black string. 
It will turn out that the 5-dimensional black hole has Petrov
type~\underline{22} and that the wrapped string has Petrov type~22
(the notation will be explained in section~\ref{s:Petrov}). Because the metrics
of type~\underline{22} form a subset of the metrics of type~22,
the smallest special class containing both limiting cases has Petrov type~$22$.
Therefore, we know that if the black hole on a cylinder is algebraically
special, it should have Petrov type~22. 

Secondly, we classify Ricci-flat metrics that
are static, have an $SO(3)$ isometry group and are in addition of Petrov
type~22. It is the latter constraint which makes the
Einstein equations much more tractable.

Thirdly, it turns out that the black hole on a cylinder does not show up in this
classification and hence is not algebraically special\footnote{One could 
suspect that the 5-dimensional black hole and the wrapped 
string are algebraically special because they both have more symmetries 
than the black hole on a cylinder; the isometry group of the space-like slices
is $SO(4)$ and $SO(3)\times U(1)$ respectively 
versus only $SO(3)$. However, there is no clear connection between metrics being
algebraically special and the existence of Killing vectors and so this argument
is not valid.}.

Our method is analogous to the
method used for constructing the rotating black hole in four dimensions. 
This metric was found by Kerr~\cite{Kerr} while 
sieving\footnote{See~\cite{Kinnersley} or~\cite{Carmeli}, page 265, for
an exhaustive list of all 4-dimensional spacetimes of Petrov type $D$.} 
through the metrics having 4-dimensional Petrov type $D$.

The rest of this article is organized as follows. In section~\ref{s:Petrov}, we
give an outline of the 5-dimensional Petrov classification. Although there 
exist good textbooks~\cite{GR} on the Petrov classification in 4 dimensions,
the generalization to 5 dimensions of this classification scheme 
has to our knowledge not appeared in the literature.
In section~\ref{s:Petrov}, we also determine the Petrov type of
the 5-dimensional Schwarzschild black hole and of 
the black string wrapping the compact circle. 
In section~\ref{s:eqs}, we give an overview of the equations one has to
solve to find static black holes that have an $SO(3)$ isometry group and have
Petrov type~22.
In section~\ref{s:22}, we classify the 
metrics that are static, have an $SO(3)$ isometry group and are of Petrov 
type~22.
In section~\ref{s:22s}, we give a classification of metrics that are static, 
have an $SO(3)$ isometry group and are of Petrov type~\underline{22}. 
In section~\ref{s:NOT}, we prove that all metrics of type~22 that 
approach the wrapped string geometry at infinity, are homogeneous along the 
circle. This
proves in particular that the black hole on a cylinder is not algebraically
special.  

We end with conclusions in section~\ref{s:conc}. Most calculations have been
referred to the appendices.
\section{Petrov classification in five dimensions}\label{s:Petrov}
The Petrov type of a metric is determined by its Weyl tensor $C_{ijkl}$. 
Originally, this classification in four dimensions was done using the
eigenspace structure and eigenvalues of the symmetric $6 \times 6$ matrix
$C_{[ij]\ [kl]}$. Imposing the Bianchi identity $C_{i [jkl]}=0$ 
is unnatural when using this two-index notation and the classification becomes
rather cumbersome. Thereafter, it was realized that the 4-dimensional 
Petrov classification is much easier if one uses spinors. One can
define a spinor equivalent to the Weyl tensor, called the Weyl spinor 
$\Psi_{abcd}$. Instead of classifying the possible Weyl tensors, 
one classifies the possible Weyl spinors. In this way, the Petrov 
classification becomes very clear and elegant.

It turns
out that in five dimensions also, it is very natural to use spinors to discuss
the Petrov classification. Therefore, we will now give a brief overview of 
spinor
conventions in five dimensions. For a more elaborate review, we refer the 
reader to
the literature, e.g.~\cite{Tools}. Our spacetimes will have Euclidean signature 
from now
on. Although our main interest lies in the construction of a \textit{Minkowski}
black hole on a cylinder, this black hole is static and can easily be
Wick-rotated. 
\subsection{Spinor conventions in five dimensions}
It is easy to find a realization of the Clifford algebra in 5 Euclidean 
dimensions
$$\{ \gamma_i,\gamma_j \} = 2 \delta_{ij}.$$
When we calculate the Petrov type of some concrete metrics later on,
we will always make the following choice
\begin{equation}\label{gammconcreet}
\g_1 = \s_1 \otimes 1,
\quad \g_2 = \s_2 \otimes 1,
\quad\g_3 = \s_3 \otimes \s_1,
\quad\g_4 = \s_3 \otimes \s_2,
\quad\g_5 = \s_3 \otimes \s_3\ .
\end{equation}
We use letters at the beginning of the alphabet 
$(a,b,c,\ldots)$ to denote spinor indices and letters in the middle
$(i,j,k,\ldots)$ to denote
spacetime indices.
If we let the $\g$-matrices act on spinors $\psi^a$ on the left, 
we see that they have the following index structure
$$\g_i \equiv (\g_i)^a_{\ b}.$$
We can then lower the spinor indices on the $\g$-matrices as follows
$$(\g_i)_{ab} = C_{ac} (\g_i)^c_{\ b},$$
where we take the charge conjugation matrix to be $C = -i \s_1\otimes \s_2.$
The matrix $(\g_i)_{ab}$ is antisymmetric. We also have
$$(\g_{ij})_{ab} = C_{ac} (\g_{ij})^c_{\ b}\qquad\mbox{where }
\g_{ij} = \frac{1}{2} [\g_i,\g_j ]. $$
This matrix is symmetric. We will use it to define the Weyl spinor in the
next section.
\subsection{The Weyl spinor}
In four dimensions, the definition of the Weyl spinor is motivated by the
isometry between $so(4)$ and $su(2)\oplus su(2)$. Similarly, in five dimensions,
we use the isomorphism between $so(5)$ and $sp(4)$ to map
representations of $so(5)$ into representations of $sp(4)$ and vice versa.
The Weyl tensor $C_{ijkl}$ of $so(5)$ is mapped into 
a tensor $\Psi_{abcd}$ of $sp(4)$, which, using the Bianchi identity, 
turns out to be completely symmetric
\begin{equation}\label{Weylt1}
so(5) \to sp(4): \YTvert\!\!\!\YTvert \to \YThor\!\YThor: \Psi_{abcd} = 
(\g_{ij})_{ab} (\g_{kl})_{cd}C^{ijkl}.
\end{equation}
We will refer to the object $\Psi_{abcd}$ as the {\it Weyl spinor}.
As an easy check for this 
correspondence, one can count the number of degrees of freedom of the 
completely symmetric tensor $\Psi_{abcd}$. 
This is 35, equal to the dimension of the Weyl 
tensor $C_{ijkl}$. It is easy to invert the relation~\eqref{Weylt1}:
\begin{equation}\label{Weylt2}
C_{ijkl} = \frac{1}{64} (\g_{ij})^{ab} (\g_{kl})^{cd}\Psi_{abcd}.
\end{equation}
\subsection[The Petrov classification]{The Petrov 
classification in five dimensions}
\begin{figure}[ht]
\begin{center}
\epsfxsize=10cm
\epsfysize=5cm
\begin{psfrags}
\psfrag{1}[][]{4}
\psfrag{2}[][]{31}
\psfrag{3}[][]{22}
\psfrag{4}[][]{211}
\psfrag{5}[][]{\underline{22}}
\psfrag{6}[][]{2\underline{11}}
\psfrag{7}[][]{1111}
\psfrag{8}[][]{\underline{11} \underline{11}}
\psfrag{9}[][]{11\underline{11}}
\psfrag{10}[][]{$\Psi = 0$}
\psfrag{11}[][]{\underline{1111}}
\psfrag{12}[][]{1\underline{111}}
\epsfbox{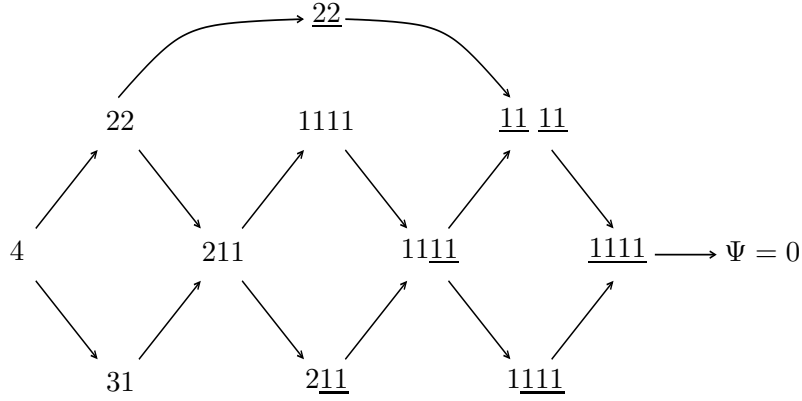}
\end{psfrags} 
\caption{The 12 different Petrov types in 5 dimensions.}
\label{fig:Ptypes}
\end{center}
\end{figure}

The fact that we have found a completely symmetric spinor equivalent of the 
Weyl tensor makes it easy to classify Weyl tensors in an invariant way. 
To this end, we
associate to the Weyl spinor $\Psi_{abcd}$ the \textit{Weyl polynomial} $\Psi$,
which is a homogeneous polynomial of degree four in four variables:
\begin{equation}\label{WP1}
\Psi = \Psi_{abcd} x^a x^b x^c x^d\ .
\end{equation}
The Petrov type of a given Weyl tensor is nothing else than the number and
multiplicity of irreducible\footnote{The Weyl polynomial has generically complex
coefficients. Hence, in this article, we will decompose this Weyl polynomial 
over $\IC$. However, the Weyl polynomial is in fact a real polynomial 
in the sense that there is a reality condition on 
the Weyl spinor $\Psi_{abcd}$ derived from the fact that the $\g$-matrices 
are Hermitian. Therefore, a more refined classification is
possible if one makes a distinction between ``real'' and ``complex''
factorizations. Furthermore, it is possible to
refine the Petrov classification by characterizing the leftover polynomials of
degree 2 by their eigenvalues, for example.}
factors of $\Psi$. This is an invariant classification scheme, because under
$O(5)$~transformations of the tetrad,
the Weyl polynomial changes by a linear transformation of 
its variables $x^a$. This transformation respects
the factorization properties of~$\Psi$.
In this way, we get 12 different
Petrov types, which are depicted in figure~\ref{fig:Ptypes}.
We use the following notation. The number denotes the degree of the
irreducible factors and underbars denote the multiplicities. For example, a Weyl
polynomial which can be factorized into two different factors, each having
degree~2, has Petrov type~22. As a second example, 
a Weyl polynomial which can be
factorized in a polynomial of degree~2 times the square of a polynomial of
degree~1 has Petrov type~2\underline{11}. The arrows between the different
Petrov types denote increasing specialization of the Weyl tensor. 
In accord with the literature on the 4-dimensional Petrov classification, 
all Weyl tensors different from type~4 are called algebraically special. 
\subsection{Examples}\label{s:examples}
\subsubsection*{five-dimensional spherical symmetric black hole}
The metric reads
\begin{equation}\label{Schw5}
ds^2 = (1+ 2 V)dt^2+(1+2 V)^{-1} dr^2 + r^2 d\O_3^2,
\end{equation}
where the function $V(r)$ is the Newton potential $V(r) = -G_5M / r^2$ and 
$d\O_3^2$ is the metric on the unit 3-sphere
$$d\O_3^2 = d\psi^2 + \sin^2\psi\left( d\theta^2 + 
\sin^2\theta d\phi^2\right).$$
If we choose the tetrad as 
\begin{alignat*}{3}
e_t &= \frac{1}{\sqrt{1+2 V}} \p_t\qquad &e_r &= \sqrt{1+2 V} \p_r\qquad
& e_{\psi} &= \frac{1}{r}\p_{\psi}\\ 
e_{\theta} &= \frac{1}{r \sin \psi}\p_{\theta}\qquad&
e_{\phi} &= \frac{1}{r \sin \psi\sin \theta}\p_{\phi},&&
\end{alignat*}
we find after some algebra the Weyl polynomial
$$\Psi = -\frac{192\ G_5 M}{r^4}\left( x t - y z \right)^2. $$
Here we have used the notation $(x^1,x^2,x^3,x^4) = (x,y,z,t)$.
Hence, the metric~\eqref{Schw5} has Petrov type \underline{22}.
\subsubsection*{Wrapped black string}
The metric of a black string wrapped around a circle of radius $R$ reads
\begin{equation}\label{Schw4S}
ds^2 = (1+ 2 V)dt^2+(1+2 V)^{-1} dr^2 + r^2 d\O_2^2+R^2 d\a^2,
\end{equation}
where $V(r)$ is the Newton potential $V(r) = -G_4 M / r$ and $d\O_2^2$ is
the metric on the unit 2-sphere.
If we choose the tetrad as 
\begin{alignat*}{3}
e_t &= \frac{1}{\sqrt{1+2 V}} \p_t\qquad &e_r &= \sqrt{1+2 V} \p_r\qquad
&e_{\a} &= \frac{1}{R}\p_{\a}\\
e_{\theta} &= \frac{1}{r}\p_{\theta}\qquad& 
e_{\phi} &= \frac{1}{r \sin\th}\p_{\phi},&&
\end{alignat*}
we find after some algebra the Weyl polynomial
$$\Psi = -\frac{96\ G_4 M}{r^3}\left( x^2 t^2 + y^2 z^2 \right).$$
The metric~\eqref{Schw4S} has Petrov type $22$ because we can factorize
$\Psi$ as
$$\Psi = -\frac{96\ G_4 M}{r^3}\left( x  t +  i y z\right)
\left( x  t - i y z\right).$$

At this point, we know the Petrov type of two limiting cases of the black hole
on a cylinder. Namely, at large distances (or, equivalently, at small
compactification scale), the metric has Petrov type~22. At small 
distances (or, equivalently, large compactification scale), 
the metric has Petrov
type~\underline{22}. Taking a look at figure~\ref{fig:Ptypes}, we see that the
black hole on a cylinder should have Petrov type~22 or type~4. In the latter
case, the metric is algebraically generic. In the next section, we give the
equations one has to solve to determine all metrics that are static, have an
$SO(3)$ isometry group and are of Petrov type~22.
\section{Metrics of type 22: equations}\label{s:eqs}
Because we
want to impose a constraint on the Weyl tensor, we will not solve the Einstein
equations in the usual way. We will not write an ansatz for the metric and 
solve the
differential equations resulting from putting the Ricci tensor to zero. Instead,
we will solve the following set of equations
for the dual tetrad~$\th$, the connection~$\o$ and the curvature~$\O$
\begin{subequations}
\label{NP}
\bq
&&d\th+\o\wedge\th=0\label{NPa}\\
&&d\o +\o \wedge \o=\O\label{NPb}\\
&&d\O + \o \wedge\O -\O\wedge\o =0\label{NPc}.
\eq
\end{subequations}
These
three equations are of course respectively the definition of zero torsion, 
the definition of
the curvature $\O$ and the Bianchi identity.
In the last two equations~(\ref{NPb}, \ref{NPc}), 
we use equation~\eqref{Weylt2} to express the Riemann 
tensor $\O$ in terms of the Weyl spinor $\Psi_{abcd}$. We can indeed use
formula~\eqref{Weylt2} for this substitution because the Riemann tensor is equal
to the Weyl spinor for spacetimes with zero Ricci tensor. Because the Weyl
spinor now enters the field equations explicitely, it is easy to impose a
condition on the Petrov type. 
\subsection{Ansatz}
The most general ansatz for the tetrad of a static spacetime with $SO(3)$
symmetry reads
\begin{equation}\label{eans}
\begin{split}
&e_t = A\ \p_t\\ 
&e_r\mbox{ and }e_{\a}:\mbox{ linear combinations of }\p_r 
\mbox{ and }\p_{\a}\\ 
&e_{\th} = \s\ \p_{\theta}\\
&e_{\phi} = \frac{\s}{\sin\th}\ \p_{\phi}
\end{split}
\end{equation}
Here, $A$ and $\s$ are functions of the coordinates $r$ and $\a$. The most
general connection consistent with the above ansatz is
\begin{equation}\label{omans}
\o=
\begin{pmatrix}
0& \m \th^1& \n \th^1&0&0\\
-\m \th^1&0 & \d \th^2+\e\th^3&\k \th^4 &\k \th^5\\
-\n \th^1 & -\d \th^2-\e\th^3&0&\l \th^4 &\l \th^5\\
0&-\k \th^4& -\l \th^4 &0&-\s \cot\th\ \th^5\\
0&-\k \th^5& -\l \th^5 &\s \cot\th\ \th^5&0\\
\end{pmatrix}.
\end{equation}
Here, $(\th^1, \ldots,\th^5)$ is the basis dual to the tetrad~\eqref{eans} and
the seven functions $(\m,\n,\d,\e,\k,\l,\s)$ are functions of the
coordinates $r$ and $\a$. Notice that although we could have expressed the
connection $\o$ in terms of the functions appearing in the tetrad~\eqref{eans},
we have not done so. Rather, as will become clear in the following, we 
determine $\o$ by solving the equations~\eqref{NPb} and~\eqref{NPc}.
Finally, a calculation shows that the most general ansatz for the Weyl spinor
$\Psi_{abcd}$ has 4 independent components.
\begin{equation}\label{Wans}
\begin{split}
&\Psi_{1114} = \Psi_{1444}=\Psi_{2223} = \Psi_{2333}=-3 \psi_1\\
&\Psi_{1123} = \Psi_{1224}=\Psi_{1334} = \Psi_{2344}=\psi_1\\
&\Psi_{1111} = \Psi_{2222}=\Psi_{3333} = \Psi_{4444}=-3 \psi_2\\
&\Psi_{1133}= \Psi_{2244}=\psi_2\\
&\Psi_{2233} = \Psi_{1144}=\psi_3\\
&\Psi_{1122} =\Psi_{3344} =\psi_4 \text{ and }
\Psi_{1234} =-\frac{1}{2}(\psi_3+\psi_4)
\end{split}
\end{equation}
Here again, the four functions $\psi_i$ are functions of the coordinates $r$ 
and 
$\a$. In some cases, equations simplify when written in terms of the 
functions $\vp_i$, defined by
\begin{equation}\label{defvp}
\begin{aligned}
\psi_1 &= -8 i \vp_4 \\
\psi_3 &=  8(\vp_2-3\vp_3)
\end{aligned}\qquad\qquad
\begin{aligned}
\psi_2 &=8 (\vp_2+\vp_3)\\
\psi_4&=8(-2\vp_1-\vp_2+\vp_3).
\end{aligned}
\end{equation}
The above linear combinations are chosen in such a way 
that the functions $\vp_i$ transform nicely under tetrad transformations 
(see appendix~\ref{A:transf}).
\subsection{Field equations}
We can now take the ans\"atze~\eqref{omans} and~\eqref{Wans} and insert them into
equations~\eqref{NP}. 
The differential equations obtained in this way are quite lengthy.
They can be found in appendix~\ref{veldvgl}.
\subsection{The Petrov condition}
In this section, we give necessary and sufficient conditions on the
components~$\vp_i$ of the Weyl spinor such that the
Weyl polynomial $\Psi$ obtained from the Weyl spinor~\eqref{Wans} has Petrov
type~$22$. 
\kaderzonderrm{The metric has Petrov type 22 if and only if one of the 
following four conditions is satisfied
\begin{equation}\label{Pcond2}
\begin{aligned}
\text{Case A: }\ &\vp_3=\vp_4=0\\
\text{Case B: }\ &\vp_1+\vp_2=0
\end{aligned}\qquad
\begin{aligned}
\mbox{Case C: }\ &\vp_1^2=\vp_3^2+\vp_4^2\\
\mbox{Case D: }\ &\vp_2^2=\vp_3^2+\vp_4^2
\end{aligned}
\end{equation}
}
The characterization is most easily derived in a frame for which $\vp_4=0$.
It is proven in appendix~\ref{A:transf} that such a frame exists. In this frame,
using~\eqref{Wans} and~\eqref{defvp} and dropping an irrelevant overall 
constant, the Weyl polynomial reads
\begin{equation}\label{psires}
\begin{split}
\Psifr &= \psi_2\left((x^2-z^2)^2+(y^2-t^2)^2\right)
-2\psi_3(x^2t^2+y^2z^2)\\
&\qquad - 2 \psi_4(x^2y^2+z^2t^2)
+4(\psi_3+\psi_4)\ xyzt\ .
\end{split}
\end{equation}
The subscript of $\Psifr$ 
is to remind the reader that we have chosen a particular
frame. We want the polynomial~\eqref{psires} to be a product of 2~factors
of degree~2. We can deduce necessary conditions for this reducibility 
in the following way. The polynomial~\eqref{psires} should still be reducible 
when we put one of the variables $x,y,z,t$ to zero. When we restrict 
$\Psifr$ to
only three variables in this manner, the algebra becomes less cumbersome and one
can show that the restricted $\Psifr$ can be factorized into 
2~factors of degree~2 if and only if one of the following four cases holds
\bq
&&\psi_2=0 
\quad\mbox{or}\quad \psi_2^2=\psi_3^2 
\quad\mbox{or}\quad \psi_2^2=\psi_4^2
 \quad\mbox{or}\quad \psi_3+\psi_4=0.\label{Pcond1}
\eq
Although these conditions were derived as necessary conditions for having 
Petrov type~$22$, it turns out that they are sufficient as well. 
In the first three cases, the polynomial $\Psifr$ is easily factorized
after writing it as a sum of squares
\begin{xalignat*}{2}
&\mbox{if }\psi_2=0,&\text{then}\quad\Psifr&=
-2\left( \psi_3 (xt-yz)^2 + \psi_4 ( xy-zt)^2\right)\\
&\mbox{if } \psi_2^2=\psi_3^2 , &\text{then}\quad \Psifr&=
\psi_2\left( x^2 \pm y^2 -z^2 \mp t^2\right)^2
 -2 (\psi_3+\psi_4) (x y - zt )^2\\
&\mbox{if } \psi_2^2=\psi_4^2 , &\text{then}\quad \Psifr&=
\psi_2\left(x^2 \pm y^2 -z^2 \mp t^2\right)^2 
-2 (\psi_3+\psi_4) (x t - yz )^2
\intertext{The factorization in the fourth case is straightforward. 
We give it for completeness.}
&\mbox{if } \psi_3+\psi_4=0, &\text{then}\quad\Psifr&=
\frac{1}{\psi_2}
\left( \psi_2 x^2 +( \psi_3+w) y^2 -\psi_2 z^2 +(\psi_4-w) t^2\right)\times\\
&&&\qquad\qquad\times
\left( \psi_2 x^2 +( \psi_3-w) y^2 -\psi_2 z^2 +(\psi_4+w) t^2\right),
\end{xalignat*}
where $w^2 = \psi_4^2-\psi_2^2$.
The conditions~\eqref{Pcond1} are expressed in a frame in which $\psi_1=0$.
Rotating this frame back to its general position leads directly to the
conditions~\eqref{Pcond2}. 
\section{Classification of metrics of type~22}\label{s:22}
Besides the five-dimensional black hole and the wrapped black string, 
both treated in section~\ref{s:examples}, we have
encountered the following metrics:
\begin{enumerate}
\item
the Kasner metric with 
exponents $p_i$
\begin{equation}\label{Kasner}
ds^2 = dt^2 + \sum_{i=1}^4 \left(\frac{t}{t_0}\right)^{2 p_i} (dx^i)^2.
\end{equation}
This metric is Ricci flat if $\sum_i p_i = \sum_i p_i^2=1$. It appears 
naturally in the classification because, as is well known, Kasner spaces 
describe the near-singularity geometry of Schwarzschild spaces.
\item
the cone on a dS-Schwarzschild space
\begin{equation}\label{kegeldSS}
ds^2 = dt^2 + 
t^2\left[\left(1-r^2+\frac{M}{r}\right) d\a^2 +
\frac{dr^2}{1-r^2+\frac{M}{r}} + r^2 d\O^2
\right]
\end{equation}
\item
a vacuum cosmological model in which space is the product of two
spheres
\begin{equation}\label{sf-sf}
ds^2 = dt^2 + a(t)^2 d\Omega^2 + b(t)^2 d\Omega^2
\end{equation}
Here, the functions $a(t)$ and $b(t)$ are determined by the vacuum 
Einstein equations. We do not know the analytical expression of the general
solution. A particular solution is the 
cone on the product of two spheres
$
ds^2 = dt^2 + t^2/3 \left( d\O^2 + d\O^2\right) 
$
\item
a vacuum cosmological model on the product of a sphere and the 
hyperbolic plane $\mathbb{H}$
\begin{equation}\label{sf-hyp}
ds^2 = dt^2 + a(t)^2 d\Omega^2 + b(t)^2 d\mathbb{H}^2
\end{equation}
\item 
a vacuum cosmological model in which space is the product of a sphere and
a plane
\begin{equation}\label{sf-vl}
\begin{split}
ds^2 = - 4 \cosh^{-6}\left(t/t_0\right)& 
\sinh^{2 - \frac{4}{\sqrt{3}}}\left(t/t_0\right)dt^2+
\sinh^{\frac{2}{\sqrt{3}}}\left(t/t_0\right)(dx^2+dy^2)\\
& +
t_0^2\cosh^{-4}\left(t/t_0\right) 
\sinh^{2 - \frac{4}{\sqrt{3}}}\left(t/t_0\right)d\O^2
\end{split}
\end{equation}
\item
the metric of another homogeneous wrapped object
\begin{equation}\label{wtube}
ds^2 = (1-2 M/r )dt^2+dr^2+R^2 (1-2 M/r)^{-1} d\a^2 + r^2(1-2M/r) d\O^2
\end{equation}
This metric has curvature singularities in $r=0$ and $r = 2 M$.
\end{enumerate}
Some of these solutions admit 3 orthogonal commuting Killing vectors and
therefore belong to the family of higher dimensional Weyl solutions that 
Reall and Emparan studied~\cite{genWeyl}. In particular, the five-dimensional 
black hole, the black string and the 
solutions~\eqref{Kasner},~\eqref{sf-vl}, and~\eqref{wtube} all belong to this 
family. 

The metrics~\eqref{Schw4S},~\eqref{sf-vl} and~\eqref{wtube} are all special
cases of the general homogeneous solution~\cite{Chodos}
\begin{equation*}
ds^2 = \left(\frac{r-m}{r+m}\right)^{2 a} dt^2
+\frac{1}{r^4} \left(r+m\right)^{2 + 2 a + 2 c}
(r-m)^{2 -2 a -2 c} \left(dr^2 + r^2 d\Omega^2\right)
+ \left(\frac{r-m}{r+m}\right)^{2 c} d\alpha^2
\end{equation*}
with $a^2+ ac + c^2 =1$. In particular, for the wrapped string~\eqref{Schw4S}, 
$a=1$ and $c=0$, for the metric~\eqref{wtube}, $a=1$ and $c=-1$ and for the 
metric~\eqref{sf-vl}, $a = c = 1/\sqrt{3}$.
The classification is then as follows\footnote{\label{brevity}For brevity, 
we have adopted
the following usage. For example, we say that case B consists of the 
five-dimensional Schwarzschild metric. It is to be understood that its two 
limits --
flat space and the Kasner space~\eqref{Kasner} with exponents 
$(-\frac{1}{2},\frac{1}{2},\frac{1}{2},\frac{1}{2})$ -- belong to case B as
well. 
}.
\kaderzonderrm{\begin{description}
\item[Case A:] the wrapped string~\eqref{Schw4S}
\item[Case B:] the five-dimensional black hole~\eqref{Schw5}
\item[Case C:]the five-dimensional black hole~\eqref{Schw5} and the 
metric~\eqref{wtube}
\item[Case D:] the 5-dimensional black hole, 
the cone on an dS-Schwarzschild space~\eqref{kegeldSS}, the
wrapped string and the cosmological models~\eqref{sf-sf},~\eqref{sf-hyp}
and~\eqref{sf-vl}
\end{description}
}
The derivation of this classification is tedious, it 
can be found in the appendices~\ref{ap:ABandC} and~\ref{ap:D}.
\section{Classification of metrics of type~\underline{22}}\label{s:22s}
From the characterization~\eqref{Pcond2}, one can easily derive the 
necessary and sufficient
conditions leading to metrics of Petrov type~\underline{22}. 
These conditions are combinations of the
conditions~\eqref{Pcond2}, which is reflected in our notation.
Details of the classification can be found in appendix~\ref{ap:22s}.
\kaderzonderrm{
The metric is of Petrov type~\underline{22} in the following three cases:
\begin{description}
\item[Case AC:] $\vp_1=\vp_3=\vp_4=0$: only flat space
\item[Case AD:] $\vp_2=\vp_3=\vp_4=0$: only flat space
\item[Case BC:] $\vp_1+\vp_2=0$ and $\vp_1^2=\vp_3^2+\vp_4^2$:
the five-dimensional black hole~\eqref{Schw5} 
\end{description}
}
\section{Study of the black hole on a cylinder}\label{s:NOT}
We now prove that the black hole on a cylinder and  -- more generally -- 
all inhomogeneous wrapped strings are not algebraically special. 

These
metrics approach the geometry of the wrapped string at infinity, hence, we see from
figure~\ref{fig:Ptypes} that they should be of Petrov type~22 if they are
algebraically special. However, in section~\ref{s:22}, we have classified all solutions 
that are static, have an $SO(3)$ isometry group and have Petrov type~22. The 
non--homogeneous wrapped string was not one of them.
Hence, we know that the black hole on a cylinder and more generally 
the non-homogeneous wrapped string are not algebraically special.
This completes the proof.
\section{Conclusions}\label{s:conc}
It is well-known that string theory suggests that spacetime has 10 or 11
dimensions. Although string theory is not sufficiently developed to make
concrete physical predictions that can be tested experimentally, it seems to be
clear that these extra dimensions should be very small for
string theory not to be in disagreement with 
reality\footnote{Recently, however, scenarios with large extra dimensions have 
been proposed~\cite{RS}.}.

It is interesting to study black holes in these higher dimensional spacetimes
for the following reasons. 
Firstly, due to the Einstein equations, black
holes will alter the geometry of these extra dimensions in their neighborhood.
It is possible that this will lead to an increase in size of these extra
dimensions close to a black hole, making these dimensions easier to detect.
Secondly, string theory predicts generically that the fundamental physical
constants like the gravitational constant, masses of elementary particles and
the fine structure constant should depend on the geometry of the 
compactification
space. Therefore, knowledge of the black hole geometry can lead to some
predictions of the behaviour of these constants near black holes.

Thirdly, Gregory and Laflamme~\cite{unstable} found that a neutral black string
wrapping a cylinder is classically unstable if its mass is sufficiently small.
Therefore, it seems natural to conjecture that these black strings will decay
into a black hole that is localized in the compact direction. However, recently,
Horowitz and Maeda~\cite{Hor1} argued that, classically, the black string would
not decay into the black hole on a cylinder but rather to a
{\it non-homogeneous} wrapped string\footnote{See~\cite{Hor3} for a nice review
and~\cite{Gubser} for numerical investigations on these non-homogeneous black
strings.}. A similar claim holds for black strings
with large masses: in~\cite{Harmark}, it was argued that black strings with
large masses have smaller entropy than the non-homogeneous wrapped strings, and
hence should decay as well.

In this article, we studied the metrics of the black hole on a cylinder and of
the non-homogeneous wrapped string using the notion of algebraically special
metrics. We have treated the Petrov classification in
five dimensions. We have given a complete classification of 
five-dimensional metrics that are Ricci-flat, static, of Petrov type~$22$ and 
have an $SO(3)$ isometry group. We used this classification to prove that the
black hole on a cylinder and the non-homogeneous wrapped string are not
algebraically special.

Some possible topics for future research are
\begin{enumerate}
\item
Classify all Ricci-flat metrics of type~$22$ without the additional 
assumptions we made.
This could lead to the discovery of new five-dimensional metrics. One knows that
the set of black holes is much richer in five dimensions than in four
dimensions. See~\cite{Gibbons} for a study of uniqueness theorems in higher
dimensions and~\cite{ring} for a five-dimensional rotating black string.
\item
The metric of a black hole in the RS-scenario~\cite{RS} that has a localized
horizon in the bulk spacetime is not known~\cite{9911043,9909205}. 
It is possible that a line of
research similar to the one presented here would lead to a construction of this
metric.
\item
It would be nice to find some other simplifying assumption yielding solvable
equations. It may be possible to find the metric of the black hole on a
cylinder by a clever choice of coordinates; for an attempt, see~\cite{Harmark}.
\end{enumerate}
\section*{Acknowledgments}
I would like to thank Yves Demasure, Stefan Vandoren, 
Antoine Van Proeyen and Walter Troost for discussions. 
I also would like to thank Pierre Savaria and Walter Troost for suggestions to
improve this text and S.~Hervik for a correction in formula~\eqref{sf-vl} in an
earlier version of this paper.
This work was supported in part by the European Commission RTN program 
HPRN-CT-2000-00131.
\appendix
\section{Tetrad transformation on the Weyl spinor}\label{A:transf}
In this appendix, we give the behaviour of the components of the Weyl spinor
under tetrad transformations. Direct use of the definition~\eqref{Weylt1},
shows that under tetrad rotations ($s = +1$) and reflections ($s = -1$) 
$$
\begin{pmatrix}
e_r\\e_{\a}
\end{pmatrix} \to 
\begin{pmatrix}
\cos\chi & \sin\chi\\
-s\sin\chi & s\cos\chi\\
\end{pmatrix}
\begin{pmatrix}
e_r\\e_{\a}
\end{pmatrix},
$$
the Weyl spinor transforms as
$$
\vp_1\to\vp_1,\quad\vp_2\to\vp_2,\quad
\begin{pmatrix}
\vp_3\\ \vp_4
\end{pmatrix} \to 
\begin{pmatrix}
\cos 2\chi & -\sin 2\chi\\
s\sin 2\chi &s\cos 2\chi\\
\end{pmatrix}
\begin{pmatrix}
\vp_3\\ \vp_4
\end{pmatrix}.
$$
From this, we have the important result that we can always use a tetrad 
rotation to put $\vp_4=0$. On top of this, we can use a reflection to specify 
the sign of $\vp_3$. 
\section{The field equations}\label{veldvgl}
In this appendix, we explicitely write down the equations 
resulting from inserting the static axi-symmetric 
ansatz~(\ref{eans}, \ref{omans}, \ref{Wans}) into the differential
equations~\eqref{NP}. We will split up these equations into three blocks. 
\subsection*{The equations from~\eqref{NPa}}
The condition of zero torsion~\eqref{NPa} yields the commutation relations 
between the tetrad vectors~\eqref{eans}. The non-zero commutators are
\begin{subequations}
\label{eq:comms}
\bq
\left[e_t,e_r\right]&=& \m e_t,
\qquad\left[e_t,e_\a\right]= \n e_t,\label{comm:et}\\
\left[e_r,e_{\th}\right]&=& \k e_{\th},\qquad [e_\a,e_{\th}]= \l e_{\th},\\
\left[e_r,e_{\phi}\right]&=& \k e_{\phi},\qquad [e_\a,e_{\phi}]= \l e_{\phi},\\
\left[e_r,e_\a\right]&=& \d e_r+\e e_\a\label{comm:erea}\\
\left[e_{\th},e_{\phi}\right]&=&- \cot\th\ \s\ e_{\phi}.
\eq
\end{subequations}
\subsection*{The equations from~\eqref{NPb}}
In the definition~\eqref{NPb} of the curvature tensor~$\O$, we use 
equation~\eqref{Weylt2} and the fact that the Ricci tensor is zero to express 
the curvature tensor in terms of the Weyl spinor $\Psi_{abcd}$. We also use
equation~\eqref{NPa} to replace terms involving $d\th^i$ that appear in $d\o$. 
In this way, we find two algebraic equations
\bq
&&\s^2-\k^2-\l^2-3\vp_1-\vp_2=0\label{eq:algs}\\
&&\k\m+\l\n + \vp_1-\vp_2=0\label{eq:alg}
\eq
\begin{subequations}
\label{eq:curvr}
and a set of first order differential equations. The ones 
involving the vector $e_r$ are
\bq
e_r(\s) &=& \s\k\\
e_r(\e) -e_{\a}(\d)&=& \d^2 +\e^2  +  \vp_1 + 3\vp_2\\
e_r(\k) &=& \k^2 - \d\l - \vp_1 - \vp_2 -2\vp_3\\
e_r(\l) &=& \k\d + \k\l + 2\vp_4\label{eq:erl}\\
e_r(\m) &=& -\n\d -\m^2-\vp_1+\vp_2-4\vp_3\label{eq:erm}\\
e_r(\n) &=& -\n\m + \d\m + 4\vp_4\label{eq:ern}
\eq
\end{subequations}
and the ones involving the vector $e_{\alpha}$ are
\begin{subequations}
\label{eq:curva}
\bq
e_{\a}(\s) &=& \s\l\\
e_{\a}(\k) &=& -\e\l +\k\l +2\vp_4\label{eq:eak}\\
e_{\a}(\l) &=&\e\k +\l^2 -\vp_1 -\vp_2 +2\vp_3\label{eq:eal}\\
e_{\a}(\m) &=& -\n\e -\n\m +4\vp_4\label{eq:eam}\\
e_{\a}(\n) &=&-\n^2 +\e\m -\vp_1 +\vp_2 +4\vp_3\label{eq:ean}
\eq
\end{subequations}
\subsection*{The equations from~\eqref{NPc}}
These are the equations resulting from the Bianchi identity. We again replace 
the curvature tensor~$\O$ by the Weyl spinor $\Psi_{abcd}$. We also use
equation~\eqref{NPa} to replace terms involving $d\th^i$ that appear in $d\O$. 
In this way, we find the following set of equations
\begin{subequations}
\label{eq:Bianchi}
\begin{align}
e_r(\vp_1) &=
     \tfrac{5}{2}\k\ \vp_1 + 
       \tfrac{1}{2}\ (\k + \m)\vp_2 + 
       \tfrac{1}{2}\ (4\k + \m)\vp_3 - 
       \tfrac{1}{2}\ (\n + 4\l)\vp_4\\ 
e_r(\vp_2) &=
     \tfrac{1}{2}\k\ \vp_1 + 
       \tfrac{1}{2}\ (5\k - 3\m)\vp_2 - 
       \tfrac{1}{2}\ (4\k + 3\m)\vp_3 + 
       \tfrac{1}{2}\ (3\n + 4\l)\vp_4\\ 
e_r(\vp_3) &=
      e_{\a}(\vp_4) + 
       \tfrac{1}{2}\k\vp_1-\tfrac{1}{2}\ (\k +\m)
	\vp_2 + (2\ \e +\k - \tfrac{1}{2}\ \m)
	\vp_3\\ 
	& \qquad+ (\tfrac{1}{2}\ \n + 2\d- \l)\vp_4\\ 
e_r(\vp_4) &=-e_{\a}(\vp_3) - 
       \tfrac{1}{2}\ \l\ \vp_1 + \tfrac{1}{2}\ (\n + \l) 
	\vp_2 + (-\tfrac{1}{2}\ \n - 
              2\d+ \l)\vp_3\\
	 &\qquad + (2\ \e +\k - \tfrac{1}{2}\ \m)\vp_4\\ 
e_{\a}(\vp_1) &=
     \tfrac{5}{2}\ \l\ \vp_1 + 
       \tfrac{1}{2}\ (\n + \l)\vp_2 - 
       \tfrac{1}{2}\ (\n + 4\l)\vp_3 - 
       \tfrac{1}{2}\ (4\k +\m)\vp_4\label{ea:vp1}\\ 
e_{\a}(\vp_2) &=
     \tfrac{1}{2}\ \l\vp_1 + 
       \tfrac{1}{2}\ (-3\n + 5\l)\vp_2 + 
       \tfrac{1}{2}\ (3\n + 4\l)\vp_3 + 
       \tfrac{1}{2}\ (4\k + 3\m)\vp_4
\end{align}
\end{subequations}
\section{Type \underline{22}: solutions of the field equations}
\label{ap:22s}
In this appendix, we classify all solutions of the field equations given in
appendix~\ref{veldvgl} that are of Petrov type~\underline{22}.
\boldmath
\subsection*{Case AC: $\vp_1=\vp_3=\vp_4=0$}
\unboldmath
From the Bianchi equations~\eqref{eq:Bianchi}, we immediately get
$(\k+\m)\vp_2=0$ and $(\l+\n)\vp_2=0$.
Hence, we have two possibilities. The first one is $\vp_2=0$. Then all 
components of the Weyl spinor are zero and we obtain flat space. The second
possibility is $\k+\m=0$ and $\l+\n=0$. Then we obtain from~\eqref{eq:algs} 
and~\eqref{eq:alg} $\s=0$. This is not a
good solution of the field equations, because in this case the
tetrad~\eqref{eans} is degenerate. Therefore, in case~AC, we obtain only flat
space.
\boldmath
\subsection*{Case AD: $\vp_2=\vp_3=\vp_4=0$}
\unboldmath
This case is similar to the previous one.
From the Bianchi equations~\eqref{eq:Bianchi}, we get $\k\vp_1 = 0$ and 
$\l\vp_1=0$. If $\vp_1=0$, we have flat space. If $\k=\l=0$, we find from
equation~\eqref{eq:alg} $\vp_1=0$, giving flat space again.
\boldmath
\subsection*{Case BC: $\vp_1+\vp_2=0$ and $\vp_1^2=\vp_3^2+\vp_4^2$}
\unboldmath
As shown in appendix~\ref{A:transf}, we can choose our tetrad in such a 
way that $\vp_4=0$ and
$\vp_3=\vp_1$. 
From the Bianchi equations~\eqref{eq:Bianchi}, we obtain 3 algebraic equations
$$\n\vp_1 =\d\vp_1=(\k-\e)\vp_1=0.$$
From these equations, we get either $\vp_1=0$ or $\n=\d=\k-\e=0$. The first
possibility leads to flat space, hence, we continue with the second possibility.
From equation~\eqref{eq:alg}, we get $\vp_1=-1/2\ \k\m$.
Inserting all this into equation~\eqref{eq:curvr} and~\eqref{eq:curva}
leads to the following equations\\
\begin{subequations}
\label{BC:curv}
\begin{minipage}{7.5cm}
\bq
&&e_r(\k) =\k^2+\k\m\\
&&e_r(\l)=\k\l\\
&&e_r(\m)= 3\k\m-\m^2
\eq
\end{minipage}
\begin{minipage}{7.5cm}
\bq
&&e_{\a}(\k) =0\label{BC:4}\\
&&e_{\a}(\l) =\k^2+\l^2 -\k\m\\
&&e_{\a}(\m) = 0\label{BC:6}
\eq
\end{minipage}
\end{subequations}
\\[2ex]
We will assume from now on that $\k\neq 0$ and $\m\neq 0$, 
because otherwise, we find flat space again. 
We choose our coordinates in such a way that $e_r = A\p_r$ and $e_\a = B\p_\a$.
It then follows from~\eqref{BC:4} and~\eqref{BC:6} that $\k$ and $\m$ do not
depend on the coordinate $\a$.
From the commutation relation~\eqref{comm:erea}, we get two differential
equations
$$
A \p_r B = \k B
\quad\mbox{and}\quad
\p_\a A= 0.
$$
Hence, A does not depend on~$r$ and we can put $A= -r \k$
by a coordinate transformation on~$r$. 
The general solution for~$B$ is $B = f(\a)/r$, where $f$ is an
arbitrary function of~$\a$. We can subsequently do a coordinate transformation
on $\a$ to put $f(\a)=1$. Hence, at this point, we have  
$e_r= -r\k \p_r$ and $e_\a = 1/r\ \p_\a$. 
The solution of the differential equations~\eqref{BC:curv} is now easily found
$$
\k=-\left(\frac{C_1}{r^4}+\frac{C_2}{r^2}\right)^{1/2},\quad
\l= - \frac{\sqrt{C_2}}{r} \cot\left(\sqrt{C_2}\a\right),\quad
\m=-\frac{C_1}{r^4}\left(\frac{C_1}{r^4}+\frac{C_2}{r^2}\right)^{-1/2}.
$$
Here $C_1$ and $C_2$ are two arbitrary integration constants. We have used a
shift on the coordinate~$\a$ to eliminate the third integration constant. 
Finally, the remaining tetrad vectors $e_t$ and $e_\th$ are 
determined from the commutation relation~\eqref{comm:et} and the 
algebraic equation~\eqref{eq:algs} respectively
$$
e_t = \left(\frac{C_1}{r^2}+C_2\right)^{-1/2}\p_t
\quad\text{and}\quad
\s=\frac{\sqrt{C_2}}{r \sin\left(\sqrt{C_2}\a\right)}\ .
$$ 
If $C_2\neq 0$, this solution is the 5-dimensional black hole~\eqref{Schw5}. 
If $C_2=0$, this solution is the Kasner
space~\eqref{Kasner} with exponents 
$(-\frac{1}{2},\frac{1}{2},\frac{1}{2},\frac{1}{2})$.
\section{Type~$\mathbf{22}$: solutions of the field equations 
in case A, B and C}\label{ap:ABandC}
In this appendix, we classify all solutions of the field equations given in
appendix~\ref{veldvgl}, subject to the conditions of case A, B and C of
~\eqref{Pcond2}.
\boldmath
\subsection*{Case A: $\vp_3=\vp_4=0$}
\unboldmath
The Bianchi identities give 
\begin{equation}\label{A:constr}
\k \vp_1 -(\k+\m)\vp_2=0
\qquad\text{and}\qquad
-\l \vp_1 +(\l+\n)\vp_2=0.
\end{equation}
Because we are not interested in the trivial case $\vp_1=\vp_2=0$, it follows
from these equations that $\k\n=\l\m$. We choose our tetrad such that $\n=0$.
At this point, it is necessary to resolve two cases: $\m=0$ and $\m\neq0$.
\begin{itemize}
\item\underline{Case A.1: $\m=0$}\\
Because $(\m,\n)$ transforms as a vector under tetrad rotations and $\m$ and
$\n$ both are zero, we still have not fixed the tetrad. 
We will fix it by choosing $\l=0$. From equation~\eqref{eq:alg} we 
obtain $\vp_1=\vp_2$. From equation~\eqref{eq:eal} then
follows that $2 \vp_1 = \e\k$. Because we are not interested in the flat case
$\vp_1=0$, it follows that $\e\neq 0$ and $\k\neq 0$. We then find from
equation~\eqref{eq:erl} that $\d=0$. We will now choose our coordinates in such a
way that $e_r = A\p_r$ and $e_\a = B\p_\a$. From equation~\eqref{eq:eak}
and~\eqref{ea:vp1}, it
follows that $\k$ and $\e$ do not depend on the coordinate $\a$. From the commutation
relation~\eqref{comm:erea}, we find two differential equations
$$
\p_\a A =0\qquad\mbox{and}\qquad A \p_r B = \e B.
$$ 
The first of these equations tells us that A depends only on the coordinate
$r$. We can then put $A= -r \k$ by a coordinate transformation on $r$. 
After these preliminary manipulations, we can start solving the
differential equations. 
The equations~\eqref{eq:curvr} yield $e_r(\e)= \e^2+2 \e\k$ and
$e_r(\k)=\k^2-\e\k$. The solution of these equations reads
$$\k = -\frac{1}{r} \left(C_1 - 2 \frac{C_2}{r}\right)^{1/2}
\qquad\mbox{and}\qquad
\e = -\frac{C_2}{r^2} \left(C_1 - 2 \frac{C_2}{r}\right)^{-1/2},
$$
where $C_1$ and $C_2$ are two integration constants.
The remaining tetrad vectors can now be determined from~\eqref{eq:comms}
and~\eqref{eq:algs}. They are
$$
e_t = \p_t,\qquad
e_\a = \left(C_1 - 2 \frac{C_2}{r}\right)^{-1/2} \p_\a,\qquad
\s= \frac{\sqrt{C_1}}{r}\ .
$$
After changing $t$ with $\a$, one sees that this solution 
is the wrapped black string~\eqref{Schw4S}. 
\item\underline{Case A.2: $\m\neq0$}\\
In this case, we have $\l=0$. From equation~\eqref{eq:ern} we find $\d=0$, and
from equations~\eqref{eq:alg},~\eqref{eq:ean} and~\eqref{eq:eal} we obtain
$$\e=-\k,\qquad
\vp_1 = -\frac{\k}{2}(\k+\m)
\qquad\text{and}\qquad
\vp_2 = \frac{\k}{2}(-\k+\m).
$$
Inserting these values into~\eqref{A:constr}, we find $\k(\k+\m)\m=0$. Hence,
$\vp_1=0$, and the treatment in case~AC of appendix~\ref{ap:22s} shows that this
leads to flat space.
\end{itemize}
\boldmath
\subsection*{Case B: $\vp_1+\vp_2=0$}
\unboldmath
We choose the tetrad in such a way that $\vp_4=0$.
From the Bianchi equations~\eqref{eq:Bianchi}, we get 
$\m(\vp_1-\vp_3)=0$ and $\n(\vp_1+\vp_3)=0$. 
Hence, there are two possibilities. The first one is $\m\neq 0$ or $\n\neq 0$.
Then we have $\vp_1=\vp_3$ or $\vp_1=-\vp_3$. Therefore the metric has
type~\underline{22}. This case is treated in case~BC of appendix~\ref{ap:22s}.
The second possibility is $\m=\n=0$. Then, it follows from~\eqref{eq:ean}
and~\eqref{eq:erm} that $\vp_1=\vp_2=\vp_3=0$. Hence, this possibility yields
flat space.
\boldmath
\subsection*{Case C: $\vp_1^2=\vp_3^2+\vp_4^2$}
\unboldmath
We choose our tetrad such that $\vp_4=0$ and $\vp_3=\vp_1$. From the Bianchi
identities~\eqref{eq:Bianchi}, we obtain two algebraic equations\\
\begin{minipage}{0.5\textwidth}
\begin{equation}\label{C:constr1}
(3 \k-2\e+\m)\vp_1+(\k+\m)\vp_2=0
\end{equation}
\end{minipage}
\begin{minipage}{0.5\textwidth}
\begin{equation}\label{C:constr2}
\text{and}\qquad
\d\vp_1=0.
\end{equation}
\end{minipage}\\[2ex]
We assume from now on that $\vp_1\neq0$, because otherwise, the discussion in
case~AC of appendix~\ref{ap:22s} shows that this leads to flat space. Hence, we
have $\d=0$. We could combine~\eqref{C:constr1} and~\eqref{eq:alg}
to obtain rather complicated expressions for $\vp_1$ and $\vp_2$. We will not do
so, however. We obtain easier expressions for $\vp_1$ and $\vp_2$ in the
following way. We let the derivative $e_r$ act on equation~\eqref{C:constr1},
then we find using equation~\eqref{C:constr1} the relation
\begin{equation}\label{C:constr3}
\vp_1+\vp_2= \frac{1}{4}(\k-\e)(\m+\e).
\end{equation}
At this point, we make a distinction between two cases, $\k=\e$ and $\k\neq\e$.
If $\k=\e$, we have $\vp_1+\vp_2=0$ and the discussion of case~BC in
appendix~\ref{ap:22s} applies. There, it was shown that this leads to the
five-dimensional black hole or a Kasner space. From now on, we assume that
$\k\neq\e$. From~\eqref{C:constr1} and~\eqref{C:constr3}, we obtain
\begin{equation}\label{C:vp}
\vp_1 = -\frac{1}{8}(\m+\e)(\k+\m)
\qquad\text{and}\qquad
\vp_2 = \frac{1}{8}(\m+\e)(3\k-2\e+\m).
\end{equation}
Because $\vp_1\neq0$, we have $\m+\e\neq0$ and $\k+\m\neq0$.
Inserting~\eqref{C:vp} into the Bianchi identities~\eqref{eq:Bianchi}, we find\\
\begin{minipage}{0.4\textwidth}
\begin{equation}\label{C:ean}
e_\a(\e) = -\e\n-\m\n
\end{equation}
\end{minipage}
\begin{minipage}{0.6\textwidth}
\begin{equation}\label{C:constr4}
\text{and}\qquad\l(\e-\k)+\n(\e+\m)=0.
\end{equation}
\end{minipage}\\[2ex]
Inserting~\eqref{C:vp} into~\eqref{eq:alg}, we find
\begin{equation}\label{C:constr5}
(\e-\k)^2-(\m-\k)^2 + 4 \l\n=0.
\end{equation}
We let the derivative $e_\a$ act on~\eqref{C:constr4}, then we find
using~\eqref{eq:curva},~\eqref{C:ean},~\eqref{C:constr4} and~\eqref{C:constr5}
the relation
\begin{equation}\label{C:constr6}
\l\n=0.
\end{equation}
Finally, because we have assumed that $\m+\e\neq0$ and $\k\neq\e$, 
we get from~\eqref{C:constr4} and~\eqref{C:constr6}
$\n=0$ and $\l=0$. From~\eqref{C:constr5},~\eqref{eq:eal} and~\eqref{eq:alg},
we then find $\e=\m$. At this point, we have squeezed the differential equations
sufficiently.
From~\eqref{eq:curvr} and~\eqref{eq:curva} we obtain
\begin{equation}\label{C:diff1}
e_r(\m) =2\k\m,\quad e_{\a}(\m) =0,\quad
e_r(\k) =\k^2+\m^2\quad\text{and}\quad e_{\a}(\k) =0.
\end{equation}
Because $\d=0$, it follows from~\eqref{comm:erea} that we can choose coordinates
such that $e_r = \p_r$ and $e_\a = B\p_\a$. The solution of~\eqref{C:diff1} is
$$\m = \frac{1}{2} \left( \frac{1}{r - C} - \frac{1}{r} \right)
\qquad\text{and}\qquad
\k = -\frac{1}{2} \left( \frac{1}{r - C} +\frac{1}{r} \right),$$
where $C$ is an integration constant. The tetrad $e_\a$ can now be determined
from~\eqref{comm:erea}
$$e_\a = \left(1 - \frac{C}{r}\right)^{1/2} \p_\a.$$
The remaining tetrad vectors can be found using~\eqref{comm:erea}
and~\eqref{eq:algs}
$$e_t = \left(1 - \frac{C}{r}\right)^{-1/2} \p_t
\qquad\text{and}\qquad
\s = \left(r^2 - C r\right)^{-1/2}.
$$
This is the metric~\eqref{wtube}.
\boldmath
\section[Type 22: solutions in case D]{Type 22: solutions in 
case D, $\vp_2^2=\vp_3^2+\vp_4^2$}
\label{ap:D}
\unboldmath
We now solve the equations~\eqref{NP} subject to the
condition $\vp_2^2=\vp_3^2+\vp_4^2$. 
We choose our tetrad such that $\vp_4=0$ and $\vp_3=-\vp_2$.
From the Bianchi
identities~\eqref{eq:Bianchi}, we obtain two algebraic equations\\
\begin{minipage}{0.5\textwidth}
\begin{equation}\label{D:constr}
\k\vp_1 + (-2 \e + 3\k)\vp_2=0
\end{equation}
\end{minipage}
\begin{minipage}{0.5\textwidth}
\begin{equation}
\text{and }\qquad
(\d-\n)\vp_2=0.
\end{equation}
\end{minipage}\\[2ex]
We will assume that $\vp_2\neq 0$, otherwise the discussion in
case~AD of appendix~\ref{ap:22s} shows that we have flat space. 
Hence, we have $\d=\n$. Other equations we obtain 
from the Bianchi identities and, using~\eqref{eq:alg}, from 
equations~\eqref{eq:curvr} read
\begin{equation}\label{D:curv}
\begin{aligned}
e_r(\vp_1)&= ( 5\e - 9 \k)\vp_2\\
e_r(\vp_2)&= ( \e + 3 \k)\vp_2
\end{aligned}\qquad\qquad
\begin{aligned}
e_r(\e) &= \e^2 + \e\m\\
e_r(\k) &= \k^2 + \k\m
\end{aligned}
\end{equation}
We now let the derivative $e_r$ act on equation~\eqref{D:constr}. 
By applying equations~\eqref{D:curv}, we find the important relation
$\e (\e-\k)\vp_2=0$. Combining this equation with~\eqref{D:constr}, it follows 
that we have to consider three cases: 
\begin{align*}
\text{D.1.} \quad &\e=\k\neq 0\quad\text{ and }\quad \vp_1+\vp_2=0\ ,\\
\text{D.2.} \quad &\e=0, \quad\k\neq 0\quad \text{ and }\quad \vp_1+3\vp_2=0\ ,\\
\text{D.3.} \quad &\e=\k=0\ .
\end{align*}
\boldmath
\subsection*{Case D.1: $\e=\k\neq 0$ and $\vp_1+\vp_2=0$}
\unboldmath
This case is treated in case~BC of appendix~\ref{ap:22s}.
\boldmath
\subsection*{Case D.2: $\e=0$, $\k\neq 0$ and $\vp_1+3\vp_2=0$}
\unboldmath
Inserting the relation $\vp_1+3\vp_2=0$ into the Bianchi equations gives 
$\l=-\n$. From equation~\eqref{eq:alg}, we get 
$\vp_2 = 1/4 (\k\m-\n^2)$. Inserting these results into
equations~\eqref{eq:curvr} and~\eqref{eq:curva} gives\\
\begin{subequations}\label{D2:er}
\begin{minipage}{7.5cm}
\begin{align}
e_r(\m) &=-3\n^2-\m^2+2\k\m\\
e_r(\n) &=0\label{D2:ern}\\
e_r(\k) &=\k^2+\k\m
\end{align}
\end{minipage}
\end{subequations}
\begin{subequations}\label{D2:ea}
\begin{minipage}{7.5cm}
\begin{align}
e_{\a}(\m) &=-\n\m\\
e_{\a}(\n) &=-\n^2\label{D2:ean}\\
e_{\a}(\k) &=-\n\k
\end{align}
\end{minipage}
\end{subequations}
\\[2ex]
We choose coordinates in such a way that $e_r = A\p_r$ and $e_\a=B\p_\a$. 
From the commutation relation~\eqref{comm:erea}, it follows that $B$ depends
only on the coordinate $\a$, hence, by a coordinate transformation, we can put
$B=1$. The general solution of~\eqref{D2:ern} and~\eqref{D2:ean} 
is $\n = 1/(\a+C_1)$, where $C_1$ is an arbitrary constant. At this point, it
is easier to make a distinction between two cases: $C_1\neq\infty$ and 
$C_1=\infty$.
\begin{itemize}
\item\underline{$C_1\neq\infty$}\\
In this case, we can do a coordinate transformation on $\a$ to put $C_1=0$. 
The general solution of~\eqref{D2:ea} then reads
$$\m = \frac{f(r)}{\a}\qquad\text{and}\qquad\k = \frac{g(r)}{\a}\ .$$
The commutation relation~\eqref{comm:erea} tells us that we can put 
$A = -\frac{r g(r)}{\a}$ by a coordinate transformation on $r$. The
equations~\eqref{D2:er} are now easily solved to yield
$$\k = - \frac{1}{\a} \left( -1 + \frac{C_3}{r^2} + \frac{2
C_2}{r^3}\right)^{1/2}
\qquad\text{and}\qquad
\m = -\frac{r^3+C_2}{r^3 \a}
\left( -1 + \frac{C_3}{r^2} + \frac{2 C_2}{r^3}\right)^{-1/2}.$$ 
Finally, 
the remaining tetrad vectors $e_t$ and $e_{\th}$ are 
determined from the commutation relation~\eqref{comm:et} and the 
algebraic equation~\eqref{eq:algs} respectively
$$e_t =  \frac{1}{\a\ r}
\left( -1 + \frac{C_3}{r^2} + \frac{2 C_2}{r^3}\right)^{-1/2} \p_t
\qquad\text{and}\qquad
\s = \frac{\sqrt{C_3}}{\a\ r}.
$$  
This the metric~\eqref{kegeldSS}.
\item\underline{$C_1=\infty$}\\
In this case, $\n=0$. The solution of~\eqref{D2:er} and~\eqref{D2:ea} can now
be found analogously. It reads
\begin{alignat*}{2}
e_r &= \left(C_3 + \frac{2C_2}{r}\right)^{1/2}\p_r 
\qquad& 
e_\a &= \p_\a\\
\k &= - \frac{1}{r} \left(C_3 + \frac{2C_2}{r}\right)^{1/2}
\qquad&
\m &= -\frac{C_2}{r^2}
\left(C_3 + \frac{2 C_2}{r}\right)^{-1/2}
\end{alignat*}
The remaining tetrad vectors $e_t$ and $e_{\th}$ are 
determined from the commutation relation~\eqref{comm:et} and the 
algebraic equation~\eqref{eq:algs} respectively
$$e_t = \left(C_3+ \frac{2 C_2}{r}\right)^{-1/2} \p_t
\qquad\text{and}\qquad
\s = \frac{\sqrt{C_3}}{r} \ .
$$
This is the wrapped black string~\eqref{Schw4S}. 
\end{itemize}
\boldmath
\subsection*{Case D.3: $\e=\k=0$}
\unboldmath
From~\eqref{eq:alg}, we get $\vp_1=\vp_2-\l\n$. From~\eqref{eq:curvr} 
and~\eqref{eq:curva} we obtain the following equations\\
\begin{subequations}\label{D3:er}
\begin{minipage}{7.5cm}
\begin{align}
e_r(\m) &=-\n^2-\m^2+\l\n+4\vp_2\label{D3:erm}\\
e_r(\n) &=0\\
e_r(\l) &=0
\end{align}
\end{minipage}
\end{subequations}
\begin{subequations}\label{D3:ea}
\begin{minipage}{7.5cm}
\begin{align}
e_{\a}(\m) &=-\n\m\label{D3:eam}\\
e_{\a}(\n) &=-\n^2+\l\n-4\vp_2\label{D3:ean}\\
e_{\a}(\l) &=\l^2+\l\n-4\vp_2\label{D3:eal}
\end{align}
\end{minipage}
\end{subequations}\\[2ex]
If $\n=0$, we get from~\eqref{D3:ean} $\vp_2=0$, then $\vp_1=0$ and we obtain
flat space. Therefore, we assume in the following that $\n\neq0$. We choose
coordinates in such a way that $e_r = A\p_r$ and $e_\a = B \p_\a$. From the
commutation relation~\eqref{comm:erea}, it follows that we can do a coordinate
transformation to make $A$ and $B$ both depend only on $\a$. Integrating the
equations~\eqref{D3:erm} and~\eqref{D3:eam}, we find
$$\m = \sqrt{C_1} \cot(\sqrt{C_1} r) A
\quad\text{and}\quad
4 \vp_2 = \n^2-\l\n-C_1 A^2,$$
where $C_1$ is an arbitrary integration constant. We now make a distinction
between the two cases $C_1=0$ and $C_1\neq0$.
\begin{itemize}
\item\underline{$C_1=0$}\\
With hindsight, by a coordinate transformation on $\a$,
we put $A= \sinh^{-\frac{1}{\sqrt{3}}}\a$. From \eqref{comm:erea}, we obtain
$B = \sqrt{3}\ \n \tanh\a$. The solution of~\eqref{D3:ean} and~\eqref{D3:eal} 
is now easily found
\begin{align*}
\n&=C_2 \cosh^4\a \sinh^{-2 + \frac{2}{\sqrt{3}}}\a\\
\l&= \frac{1}{\sqrt{3}}\left( 6 \tanh^2\a+(-3 + 2 \sqrt{3})\right) \n
\end{align*}
Here, we have used some freedom left over in the choice of coordinates to 
eliminate
one of the two integration constants. The remaining tetrad vectors can be found
from~\eqref{eq:algs} and~\eqref{comm:et}. All in all, we have
\begin{xalignat*}{2}
e_t&= \frac{1}{r} \sinh^{-\frac{1}{\sqrt{3}}}\a\  \p_t&\qquad
e_\a&= C_2 \sqrt{3}\ \cosh^3\a\ \sinh^{-1+\frac{2}{\sqrt{3}}}\a\ \p_\a\\
e_r&= \sinh^{-\frac{1}{\sqrt{3}}}\a\ \p_r&\qquad 
\s&= i C_2 2 \sqrt{3}\ \cosh^2\a\ \sinh^{-1+\frac{2}{\sqrt{3}}}\a
\end{xalignat*}
One can see by changing some notation, that this solution is the
metric~\eqref{sf-vl}.
\item\underline{$C_1\neq0$}\\
From~\eqref{eq:algs} and~\eqref{eq:comms}, it follows that the metric can
be written as
$$
ds^2 = \frac{1}{A(\a)^2} \left( dr^2 + \sin^2(\sqrt{C_1} r) dt^2 \right)
+ \frac{1}{B(\a)^2} d\a^2 + \frac{1}{\s(\a)^2}d\O^2.
$$
This metric belongs to the class~\eqref{sf-sf} if $C_1 >0$, and to the 
class~\eqref{sf-hyp} if $C_1 <0$. We have not been able to find an analytical
form of the general solution of~\eqref{D3:er} and~\eqref{D3:ea}.
\end{itemize}

\end{document}